\documentclass[prl,twocolumn,superscriptaddress]{revtex4-1}

\usepackage{amsmath}
\usepackage{amssymb}
\usepackage{xspace}
\usepackage{graphicx}
\usepackage{grffile}
\usepackage{nicefrac}
\usepackage{stmaryrd}
\usepackage{adjustbox}
\usepackage{color}

\DeclareMathOperator{\Tr}{Tr}

\begin{document}

\title{The role of non-spherical double counting in DFT+DMFT:
total energy and structural optimization of pnictide superconductors}

\author{Oleg Kristanovski}
\affiliation{I. Institut f{\"u}r Theoretische Physik, Universit{\"a}t Hamburg,
D-20355 Hamburg, Germany}
\author{Alexander B. Shick}
\affiliation{Institute of Physics, Czech Academy of Sciences,
Na Slovance 2, 182 21 Prague, Czech Republic}
\author{Frank Lechermann}
\affiliation{I. Institut f{\"u}r Theoretische Physik, Universit{\"a}t Hamburg,
D-20355 Hamburg, Germany}
\author{Alexander I. Lichtenstein}
\affiliation{I. Institut f{\"u}r Theoretische Physik, Universit{\"a}t Hamburg,
D-20355 Hamburg, Germany}


\begin{abstract}
A simple scheme for avoiding non-spherical double counting in the combination of density
functional theory with dynamical mean-field theory (DFT+DMFT)is developed. It is applied to
total-energy calculations and structural optimization of the pnictide superconductor LaFeAsO.
The results are compared to a recently proposed "exact" double-counting formulation. Both
schemes bring the optimized Fe-As interatomic distance close to the experimental value.
This resolves the long standing controversy between DFT+DMFT and experiment for the structural
optimization of LaFeAsO.
\end{abstract}

\maketitle

\section{Introduction}

Recent progress in realistic electronic structure calculations of correlated materials is
based on a combination of the density functional theory (DFT)  with the dynamical mean-field
theory (DMFT) ~\cite{kotliar_review,vollh2017}. The DFT+DMFT approach opens unique possibilities
to investigate electronic and structural properties of solids with partially-filled
$d$- and $f$-electron shells. The main reason for this success is based on the optimal nature
of the local self-energy scheme in the DMFT method~\cite{geo96}, and on the development of
efficient multi-orbital impurity solver within continuous-time quantum Monte-Carlo (CT-QMC)
schemes~\cite{CTQMC_rev}.

The ability to treat the non-spherical part of local Coulomb interactions exactly in the
CT-QMC scheme brings an additional aspect to the so-called double-counting correction in the
DFT+DMFT approach, which is commonly used to account for Coulomb interactions already treated
on the DFT level.  In a standard DFT+DMFT scheme~\cite{kotliar_review} the double-counting
corrections are spherical and are designed to repair only the average Coulomb interactions
either in the DFT or the DMFT parts.
Usually, as in the static mean-field like DFT+U scheme, the double-counting correction consists
of a subtraction of an average Coulomb interaction, that is taken either in the limits of
itinerant or localised electrons ~\cite{LDA+U_rev}.  With the full-potential DFT approach for
different structural calculations, this will work only for a strictly spherical type of
Hubbard $U$-corrections~\cite{du98}.

Already in the non-spherical rotationally-invariant DFT+U investigations of orbital ordering
and structural instability in the KCuF$_3$ perovskite~\cite{lic95}, care was taken to
avoid full potential contributions of $d$-electrons in the DFT-part. Applications of
the rotationally invariant DFT+U scheme to calculations of the complex crystal structure of
cuprates~\cite{FPLO} and magnetic-anisotropy problems~\cite{FPLO_anis} show
importance of accurate treatments of the double-counting corrections.

Another way to solve the problem of the proper DFT+DMFT interface is related to transfer
the double counting corrections to the DFT part, in order to substract the part of
exchange-correlations energy related with $d$- or $f$- electrons~\cite{Nekrasov_LDA'+DMFT}.
Recently, a so-called "exact" double-counting correction to the DFT+DMFT scheme employs a
similar idea to subtract the exchange-correlations term that correspond to local
Yukawa-like short-range interaction~\cite{hau15}. It is not clear which scheme is
more appropriate for different classes of materials. For instance, a successful application
of DFT+DMFT to the  complicated problem of the anisotropic Fermi surface of
Sr$_2$RuO$_4$~\cite{Pavarini_DC16} used a standard mean-field like double-counting
correction for the itinerant limit.

In this communication, we introduce a proper double-counting scheme for the atomic limit
and make comparison with the recent "exact" scheme ~\cite{hau15}. As a test case we
choose the problem of structurally optimizing the Fe-As distance in the pnictide
superconductor LaFeAsO.

\section{Methodology}

In a practical implementation, the total energy of the charge self-consistent DFT+DMFT
reads~\cite{pou07,dimarco2009},
\begin{eqnarray}
E^{\rm DFT+DMFT}&=&E^{\rm DFT}[\rho^{\rm DMFT}(\boldsymbol {r})]+
\sum_{\boldsymbol k}\sum_{\nu}
\epsilon_{\boldsymbol {k}\nu}\Delta N^{(\boldsymbol {k})}_{\nu\nu}\nonumber\\
&+&\langle\hat H_{\rm int}\rangle-E_{\rm dc} \; ,
\label{eq:1}
\end{eqnarray}
where  $E^{\rm DFT}$ is a standard DFT functional acting on the DMFT charge density
$\rho^{DMFT}$, $\epsilon_{\boldsymbol {k}\nu}$ are the Kohn-Sham (KS) energy eigenvalues,
$\Delta N^{(\boldsymbol {k})}$ is the KS occupation matrix correction due to the
DMFT self-energy~\cite{lec06}, $ <\hat H^{int}>$ is an expectation value of the
Coulomb vertex, and $E_{\rm dc}$ marks the double-counting correction. Eq.~\ref{eq:1}
assumes the use of the Bloch basis in which the kinetic energy operator is diagonal
in a basis of the Kohn-Sham eigenstates.

The double-counting correction $E_{\rm dc}$ in eq.~\ref{eq:1} accounts approximately
for the mean-field value of the electron-electron interaction, already included in
$E^{\rm DFT}$. Until recently, there was no precise solution of for the double counting
when utilizing conventional DFT implemented in the local-density or generalized-gradient
approximations (LDA or GGA).
Most commonly used double counting correction forms in the DFT+DMFT scheme are the
so-called''fully localized (or atomic-like) limit" (FLL)~\cite{sol94,lic95}
\begin{equation}
E_{\rm dc}^{\rm (FLL)}
= \frac{U}{2}N(N-1) - \frac{J}{2}\sum_{\sigma} N_{\sigma}(N_{\sigma}-1)\;,
\end{equation}
or, the "around mean field" (AMF) scheme~\cite{ani91,sol98},
\begin{equation}
E_{\rm dc}^{\rm (AMF)}
= U n_{\uparrow}n_{\downarrow} +
\frac{1}{2}\left( n_{\uparrow}^2 +n_{\downarrow}^2 \right)\frac{2l}{2l+1}\left(U-J\right)\;,
\end{equation}
where $n_{\sigma} = \Tr[n_{m \sigma, m^{\prime} \sigma}]$,
$n= n_{\uparrow}+n_{\downarrow}$ is the total $d$(or $f$) on-site occupation, and
$U$ and $J$ are the intra-atomic Coulomb repulsion and exchange parameter,
respectively~\cite{ani911}.
This $E_{\rm dc}$ stems from a spherically-symmetric treatment, while the DFT part of the
Hartree and the exchange-correlation energies,
\begin{equation}
\label{eq:6}
E_{\rm H} + E_{\rm XC} =
\frac{1}{2} \int d \vec{r} d \vec{r}'
\frac{\rho(\vec{r}) \rho(\vec{r}')}{|\vec{r} - \vec{r'}|} + \int d
\vec{r} \rho(\vec{r}) \epsilon_{xc}(\rho(\vec{r})),
\end{equation}
remain accounted together with the non-spherical contributions into the DFT+DMFT energy
functional eq.~\ref{eq:1} (for simplicity, we write everything in terms of a charge density
only, while the inclusion of the spin is straightforward).

One way to exclude this "non-spherical" double counting is to keep only the
spherically-symmetric contributions in the $\langle\hat H_{int}\rangle$ term of
eq.~\ref{eq:1}~\cite{du98}. But this is not what one truly aims for, the DFT+DMFT induced
enhancement of the orbital polarization beyond DFT will be lost.

Alternatively, the non-spherical contributions entering the DFT part of the Hartree and
the exchange-correlation energies from the $d$ (or $f$) states can be excluded
in a simple way, similar to what was proposed earlier in DFT+U~\cite{shi04},
and DFT+HIA~\cite{Shi09,del17}  implementations
of the full-potential linearized augmented plane wave (FLAPW) method~\cite{wim81}.

In this work, we make use of the mixed-basis pseudopotential method
(MBPP)~\cite{lou79,fu83,els90,mbpp_code}, and expand the KS wave function
 for Bloch vector $\boldsymbol{k}$ and band $\nu$ into plane waves (pw) and localized
functions (lf),
\begin{equation}
\psi_{\boldsymbol{k}\nu}(\boldsymbol{r})=\frac{1}{\sqrt{\Omega_c}}\sum_{\boldsymbol{G}}
\psi_{\boldsymbol{G}}^{\boldsymbol{k}\nu}\,{e}^{i(\boldsymbol{k}+
\boldsymbol{G})\cdot\boldsymbol{r}}+
\sum_{\gamma lm}\beta_{\gamma lm}^{\boldsymbol{k}\nu}\,
\phi_{\gamma lm}^{\boldsymbol{k}}(\boldsymbol{r})\quad,
\label{eq:2}
\end{equation}
where $\Omega_c$ is the unit-cell volume, $\boldsymbol{G}$ a reciprocal-lattice vector,
$\gamma$ labels an atom in the unit cell and $lm$ are the usual angular-momentum quantum
numbers. The localized functions are given by
\begin{eqnarray}
\phi_{\gamma lm}(\boldsymbol{r})&=& i^l\,g_{\gamma l}(r)\,K_{lm}(\hat{\boldsymbol{r}})
\qquad,\quad \nonumber\\
\phi_{\gamma lm}^{\boldsymbol{k}}(\boldsymbol{r})&=&
\sum_{\boldsymbol{T}}{e}^{i\boldsymbol{k}\cdot(\boldsymbol{T}+
\boldsymbol{R}_{\gamma})}\phi_{\gamma lm}
(\boldsymbol{r}-\boldsymbol{T}-\boldsymbol{R}_\gamma)\quad,
\label{eq:2}
\end{eqnarray}
whereby $g$ is a radial function, and $K$ is a cubic harmonic.

Accordingly, the MBPP electronic charge density $\rho(\boldsymbol{r})$ consists of
three terms, i.e.
\begin{equation}
\rho(\boldsymbol{r})=\sum_{\boldsymbol{k}\nu}
f_{\boldsymbol{k}\nu}|\psi_{\boldsymbol{k}\nu}(\boldsymbol{r})|^2=
\rho^{pw, pw}(\boldsymbol{r})+\rho^{pw, lf}(\boldsymbol{r})+
\rho^{lf, lf}(\boldsymbol{r})\quad.
\label{eq:3}
\end{equation}
For our concerns, the purely-local third term $\rho^{lf,lf}$ is of key interest.
It is written as
\begin{eqnarray}
\rho^{lf,lf}(\boldsymbol{r})&=&\sum_{\boldsymbol{k}\nu}f_{\boldsymbol{k}\nu}\left|
\sum_{\gamma lm}\beta_{\gamma lm}^{\boldsymbol{k}\nu}\,
\phi_{\gamma lm}^{\boldsymbol{k}}(\boldsymbol{r})\right|^2 \nonumber\\
&=& \sum_{\boldsymbol{T},\gamma lm}\rho^{lf,lf}_{\gamma lm}(r)\,K_{lm}(\hat{\boldsymbol{r}})\quad,
\label{eq:4}
\end{eqnarray}
with $\boldsymbol{r}'=\boldsymbol{r}-\boldsymbol{T}-\boldsymbol{R}_\gamma$, and
hence can be understood as an expansion into the cubic harmonics on each site
$\boldsymbol{R}_{\gamma}$.

We spherically average the purely local term $\rho^{lf,lf}$ in eq.~\ref{eq:4} for
those states which are corrected by DMFT (with $l=2$ for the $d$ states, and $l=3$ for
the $f$ states).  Thus, the local non-spherical parts in $\rho^{lf,lf}_\gamma$
vanish on each site $\boldsymbol{R}_{\gamma}$, and do not contribute to
the DFT part of the Hartree and the exchange-correlation energies eq.~\ref{eq:6}.
It removes the non-spherical double counting in the DFT+DMFT total energy
charge self-consistent calculations.

Until recently, there was no exact solution of the double-counting problem.
Density funtional theory does not have a diagrammatic representation that would provide an
explicit identification of corresponding many-body interaction terms. Also, it is not clear
 how to solve the Hubbard model by DFT. The FLL/AMF forms of $E_{\rm dc}$, which were
discussed above, are derived in some static approximations to the Hubbard interaction term.
The "physical" arguments prevailed in the choice of $E_{\rm dc}$.

In Ref.~\cite{hau15}, K. Haule proposed a new ''exact" form of the double-counting
correction making use of the Luttinger-Ward functional representation for both DFT and DMFT.
This $E_{\rm dc}$ was applied  to a number of correlated solids, and good agreement
between the theory and experiment was achieved. Importantly, the $E_{\rm dc}$ of
Ref.~\cite{hau15} is free from the non-spherical double counting, and no additional
correction is required. Therefore, we applied it to the structural optimisation of the
pnictide superconductor LaFeAsO.

\section{Results}

\begin{figure}[t]
\includegraphics*[width=8.5cm]{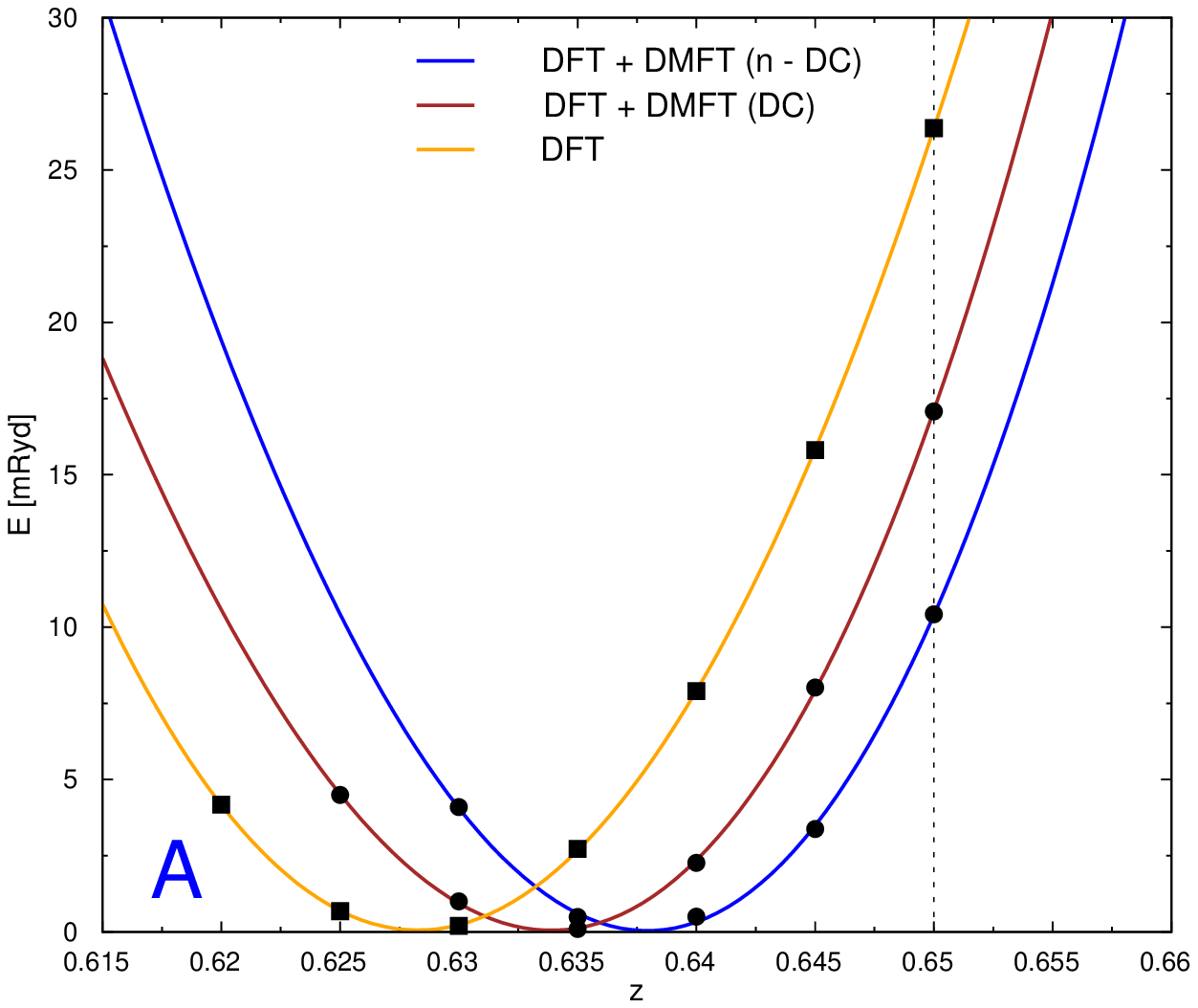}
\includegraphics*[width=8.5cm]{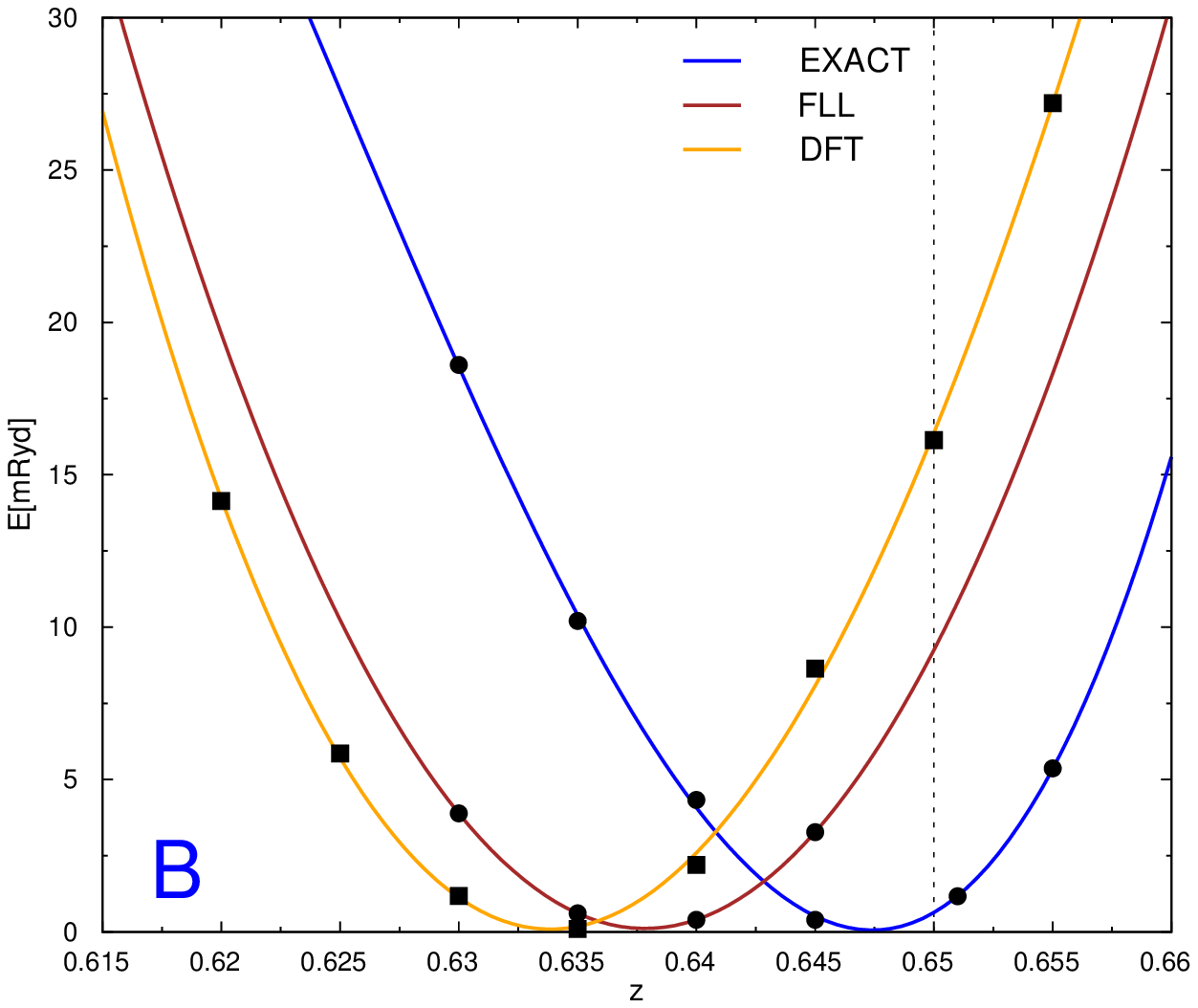}
\caption{(color online) The relative total energy of LaFeAsO as a function of the
As height, expressed by the $z$ parameter, in the unit cell, A: MBPP based,
B: Wien2K based).
Dashed line marks the experimental $z$ position of the As atom.}\label{fig:toten}
\end{figure}

\begin{table}[b]
\begin{ruledtabular}
\begin{tabular}{lc|lc}
MBPP           & $z$     & Wien2K & $z$      \\
\hline
DFT            & 0.628 & DFT          & 0.634  \\
DMFT(FLL)      & 0.633 & DFT(FLL)     & 0.638  \\
DMFT(FLL+NSPH) & 0.638 & DFT("exact")  & 0.648 \\
\end{tabular}
\end{ruledtabular}
\label{tab1}
\caption{Comparison of the As atom z position calclulated with different methods and codes.}
\end{table}

Electronic structure theory of the high-temperature superconductor LaFeAsO occupies a
fundamental place in condensed matter physics and material science. The calculations are
often performed either within the DFT or DFT+DMFT. Both approaches fall short in the correct
description of the equilibrium crystal structure. When the paramagnetic high-temperature phase
is modelled by a non-magnetic DFT calculation, a too short Fe-As distance, governed by
the internal unit-cell parameter $z$, is obtained. The latters has a drastic influence on the
low-energy electronic structure~\cite{ma}, and very precise electron-electron correlation
effects need to be tackled. Charge self-consistency within DFT+DMFT becomes important and
based thereon, previous studies~\cite{aic09,aic11} indeed improved upon pure nonmagnetic DFT
calculations. However still, those correlated electronic structure results remain ambiguous,
and depend on the choice of the double counting correction (FLL or AMF).

We overtook the values of Hubbard-$U=2.7$\,eV and Hund's exchange of $J=0.8$\,eV for the
local Coulomb interactions from Ref.~\cite{aic09}, which were calculated within
the constrained random-phase approximation (cRPA)~\cite{miy10,miy08}. The inverse
temperature is set to $\beta=40$\,eV$^{-1}$, that corresponds to room temperature $T=290$\,K.
For the solution of the quantum impurity problem we apply the cthyb-QMC method~\cite{wer06}.
The multiorbital Hubbard Hamiltonian of Slater-Kanamori form, parametrized by Hubbard-$U$
and Hund's exchange $J$ is applied to the respective full five-orbital $3d$ manifold.

We performed the total-energy electronic structure DFT and DFT+DMFT calculations of
LaFeAsO making use of the MBPP method~\cite{gri12} and compared them with Wien2K
calculations~\cite{hau10}. The FLL form of the double-counting correction was used, as well
as the "exact"  double-counting from Ref.~\cite{hau15}.  The total energy within the MBPP-based
schemes are shown in Fig.~\ref{fig:toten}A, and are compared with Wien2K-based results
shown in Fig.~\ref{fig:toten}B. The values of the As atom $z$ parameter that correspond to
the total-energy minima are given in Tab.~\ref{tab1}.

Both MBPP and Wien2K DFT calculations yield for the As atom $z$ parameter values substantially
smaller (see Tab.~\ref{tab1}) than the experimental value of $z=0.651$~\cite{ma}. Inclusion
of correlation effects by DFT+DMFT(FLL) without the non-spherical double-counting correction
has visible effect on the total energy, and improves the As atom $z$ parameter over the DFT
results. This is in agreement with the previous Wien2K-based DFT+DMFT calculations~\cite{aic09}.
Still the difference $\Delta z=0.013$ between the experimental and theoretical values
remains unresolved.

Note that within identical setting, DFT and DFT+DMFT calculations based on the
pseudopotential (MBPP) method produce $z$ parameter values by $\approx 0.005$ smaller
than corresponding all-electron (Wien2K) calculations. Nevertheless, the difference between
DFT and DFT+DMFT results obtained with MBPP is the same as from the Wien2K calculations, and
illustrates the important role of electron correlation effects.

It was proposed in Ref.~\cite{aic11} that futher improvement of the
value of $z=0.643$ can be achieved by switching to the AMF form of $E_{\rm dc}$, suggesting
partial delocalization of the $3d$ states in metallic LaFeAsO.
However, no non-spherical double-counting corrections were used in these calculations.
As it follows from our MBPP results shown in Fig.~\ref{fig:toten}A and Tab.~\ref{tab1},
avoiding non-sphericity in the FLL double counting leads to an increase of the $z$ parameter
to $z=0.638$ over the $z=0.633$ FLL result with no non-spherical double-counting correction.
Taking into account that MBPP yields slightly smaller values for the $z$ parameter, the
proposed double counting correction brings the total energy minimum into close proximity of
the experimental data.

Finally, we performed calculations with the "exact" double counting implementation~\cite{hau15},
and obtained the total-energy minimum at $z=0.648$ (see Fig.~\ref{fig:toten}B and
Tab.~\ref{tab1}.), now shifted close to the experimental value~\cite{ma}. This form is free
from the non-spherical double counting. This supports our finding that the source of
discrepancy between experiment and DFT+DMFT is not a form of the double-counting due
to the metallic character of LaFeAsO, as suggested by Ref.~\cite{aic11}, but because of
the non-spherical double counting in DFT and DMFT parts of DFT+DMFT.

To conclude we developed a simple scheme for avoiding non-spherical double counting in
DFT+DMFT and compared with the ``exact'' double counting scheme ~\cite{hau15}. As a proof
of principles, the results show a similar shift of the Fe-As distance and bring results of
DFT+DMFT closer to experiments.
We think that the standard double-counting scheme in the atomic limit will be useful for
strongly correlated $d$- and $f$ systems with anisotropic Coulomb interaction close to
insulating states.

\section{Acknowledgments}
We thank  Kristjan Haule and Eva Pavarini for helpful discussions.
Financial support was provided by
the Deutsche Forschungsgemeinschaft (DFG) Grant No. DFG LI 1413/8-1
and grant No. DFG LE 2446/4-1, as well as the Czech Science Foundation
(GACR) Grant No.~15-05872J.
Computations were performed at the University of Hamburg
and at the North-German Supercomputing Alliance (HLRN) under Grant No. hhp00040 .

\bibliography{bibextra}

\end{document}